\documentclass{epl}
\usepackage{amssymb}

\title{Gases, liquids and crystals in granular segregation}
\author{P.M. Reis\inst{1}\thanks{E-mail: \email{pedro@reynolds.ph.man.ac.uk}}, G. Ehrhardt
\inst{2}, A. Stephenson\inst{2} and T. Mullin\inst{1}}
\shortauthor{P.M. Reis \emph{et. al.}}
 \institute{
  \inst{1} Manchester Center for Nonlinear Dynamics,\\
  \inst{2} Theoretical Physics Group,\\
        Department of Physics and Astronomy, University of Manchester,\\
        Oxford Road, Manchester, M13 9PL, UK\\
}


\begin{document}

\maketitle

\begin{abstract}
We report the results of an experimental investigation of
segregation in a binary  mixture of dry particles subjected to
horizontal oscillatory excitation. The thin layer of particles was
driven by the stick--slip frictional interaction with the surface
of a horizontal tray.  As the packing fraction of the mixture was
increased the evolution of distinct phases was observed. We
identified them as a binary gas, segregation liquid and
segregation crystal and provide both microscopic and macroscopic
measures to identify their properties. Finally, we draw some
analogies between segregation in our granular system and
self-assembly in binary colloidal mixtures.

\end{abstract}

In segregation, excitation via flow or shaking can,
counter-intuitively, cause an initially homogeneous mixture of
grains to de-mix \cite{mullin:2002,shinbrot:2000}. Despite decades
of research into the phenomena \cite{williams:1976}, a predictive
model of the processes involved has yet to emerge. More recently,
segregation has received considerable attention from the physics
community as an example of a challenging far from equilibrium
system \cite{shinbrot:2000}. Although a variety of geometries has
been explored \cite{mullin:2000,breu:2003,choo:1997} and many
mechanisms proposed \cite{ottino:2000,aranson:1999}, an
understanding of the fundamental principles involved remains
incomplete. This is surprising considering that a better
understanding would have a major economic impact in the
pharmaceutical, chemical processing and civil engineering
industries. For the class of quasi-2D binary granular systems a
qualitative segregation mechanism has been suggested
\cite{shinbrot:2001,duran:1999,aumaitre:2001} using the idea of
\emph{excluded volume depletion} as in colloidal systems and
binary alloys \cite{hill:1994}. This is in the spirit of the
physics of complex systems where attempts have been made to unify
descriptions of granular materials, colloids, gels and foams
\cite{liu:1998,trappe:2001}. We discuss the relevance of these
ideas to our observations in some concluding remarks.

\begin{figure}
\begin{center}
\includegraphics[width=9cm]{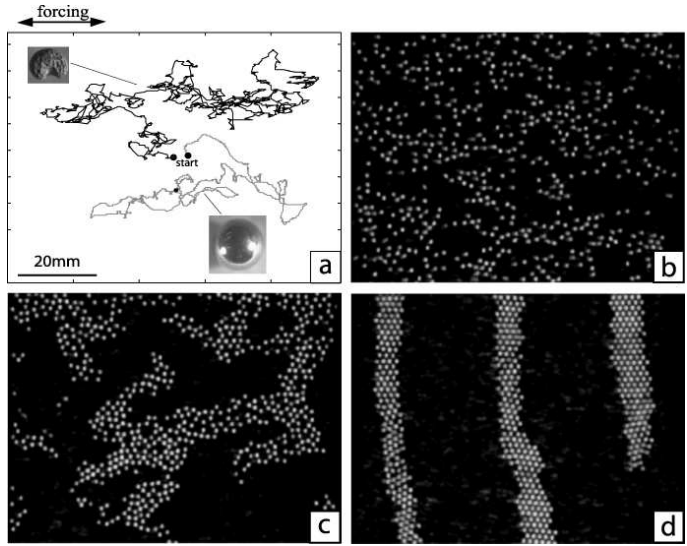}
\caption{(a) Stochastic trajectories taken over a period of 1 min
for a single particle on the tray.  The black and grey
trajectories (poppy seed and phosphor-bronze sphere, respectively)
show two quasi-random walks on the oscillating surface. The
distributions of step lengths are exponential (poppy seeds) and
gaussian (spheres) \cite{reis:2003}. The three photographs
correspond to states within the various phases: (b) binary gas
($C=0.516$), (c) segregation liquid ($C=0.729$) and (d)
segregation crystal ($C=1.071$). The binary layer was started from
mixed initial conditions in each case and frames were taken after
3min of shaking so that a (macroscopic) quasi-steady state was
achieved. Note that only the phosphor-bronze spheres are visible.}
\end{center}\end{figure}

Here we report the results of an experimental study of granular
segregation in a thin layer of a mixture of two particle types
which lie on the surface of a horizontally oscillated tray. The
tray was $180mm$ in the direction of the oscillation and $90mm$
wide and the shallow binary granular mixture was approximately one
particle deep. The particles used were monodisperse
phosphor-bronze particles which were smooth high-precision spheres
of diameter $1.5mm$ and density $8.8gcm^{-3}$ and poppy seeds
which were rough non-spherical particles with average diameter
$1.07mm$ and density $0.2gcm^{-3}$. The poppy seeds had a measured
polydispersity level of $17\%$. The size ratio for the two types
of particles used was $q\sim1.07/1.5=0.71$. The vibration was
sinusoidal to within $0.1\%$, unidirectional with fixed frequency
$12Hz$ and amplitude $1.75\pm0.01mm$ and was monitored using
accelerometers. A reasonable approximation to a two-dimensional
system was achieved, as only a small degree of overlap of the
poppy seeds occurred. This arrangement enabled us to image the
entire system using a CCD digital camera and perform particle
tracking of the phosphor-bronze spheres. High image resolution was
required for the microscopic measures and, for these, we focused
on a central $(76 \times62) mm$ area of the layer.

Two further features of our set up were that the horizontal
geometry minimized compaction effects due to gravity and the
particles were always in contact with the excitation provided by
the oscillatory tray. While the forcing was strictly sinusoidal,
both types of particles were driven by a stick-and-slip
interaction with the driving surface, resulting in anisotropic
random motion in two dimensions. Evidence for this can be seen in
the single particle trajectories shown in Fig. 1a. In addition,
the statistics of the motion were anisotropic with a preference
for motion in the direction of the forcing. Note that the two
types of particles respond to the driving differently, which
presumably arises from a combination of the differences in mass,
shape and surface properties. Details of the statistical analysis
of the particle trajectories will be presented elsewhere
\cite{reis:2003}. Clearly, the collective behaviour when many
particles are present will be modified. A simple test was carried
out by observing the motion of a coloured line within a monolayer
of non-spherical particles. This provided evidence for diffusion,
faster in the direction of the applied forcing.

The parameter space of the system is large and we have chosen to
explore the effect of the total filling fraction of the mixture,
which was previously shown to be a key control parameter
\cite{reis:2002}. We denote this by \emph{layer compacity} defined
as, $C=(N_{1}A_{1}+N_{2}A_{2})/(xy)$, where $N_{1}$ and $N_{2}$
are the numbers of poppy seeds and phosphor-bronze spheres in the
layer, $A_{1}=(0.90\pm0.15)mm^2$ and $A_{2}=(1.77\pm0.06)mm^2$ are
the two dimensional projected areas of the respective individual
particles and $x$ and $y$ are the longitudinal and transverse
dimensions of the tray. The uncertainty associated with the
measurements of $C$ is approximately $8\%$. Note that our
criterion for a quasi-two dimensional system is that the
phosphor-bronze spheres remain in the monolayer regime but, at the
highest compacities, the poppy seeds can overlap up to a maximum
layer height of $1.5mm$, i.e. the diameter of the larger spheres.
Hence, our compacities can be higher than the maximum packing
value in two dimensions. The compacity was varied by incrementally
increasing $N_1$, while keeping $N_2=1596$ constant. Experiments
showed that the phase behaviour is robust over a range of
phosphor-bronze spheres.

Analysis of the segregation domains, in previous work
\cite{reis:2002}, suggested the existence of a single continuous
phase transition of granular segregation such that there is a
critical compacity point, $C_c$, below which segregation does not
occur. Here we present new evidence that a connection with
equilibrium phase transitions may be deeper by uncovering three
qualitatively distinct segregation phases which can be related to
behavior observed in binary colloidal systems.

In Fig. 1(b,c,d) we present snapshots of three states  of the
granular layer, at different values of $C$, representative of the
three distinct phases of granular segregation.  These were
obtained in experimental runs which were consistently started from
homogeneously mixed initial conditions and only the compacity was
varied between each realization. At low $C$, Fig 1b, a
`\emph{binary gas}' was found, i.e. this was a collisional regime
where segregation did not occur. At intermediate $C$, Fig. 1c,
clustering of the larger particles occurred and mobile segregation
domains formed. The movement of the clusters across the granular
layer was reminiscent of oil drops on water as they flowed, merged
and split; the motion of the particles within the clusters was
highly agitated. We denote this by `\emph{segregation liquid}'. At
high $C$ well defined immobile stripes formed perpendicular to the
direction of forcing. Within these segregation clusters, the
spheres crystallized into an ordered hexagonal lattice, Fig 1d. We
denote this phase by `\emph{segregation crystal}'. We stress again
that these segregated states resulted from a self organization of
the granular layer, having started from a homogeneous mixture.

In order to classify these phases  we measured three quantities:
one 'macroscopic' (i.e. at the domain level) and two 'microscopic'
(i.e. at the particle level and determined by the positions of the
large spheres). Macroscopically, we studied the fluctuations of
the average width for the domains of phosphor-bronze spheres. At
the microscopic level we measured both the particle radial
distribution function and the local density for the spheres.

In Fig. 2a we present 10min long time-series of the average width
of the domains, $\phi$ as measured in \cite{reis:2002}, for three
values of the compacity which are representative of each of the
segregation phases. In the binary gas phase, at low compacities,
$\phi$ shows no evolution since the binary layer remains in a
mixed state. At high compacities, $\phi$ exhibits a fast initial
segregation growth which is followed by saturation as stable
crystalline segregation domains form. At intermediate compacities,
in the segregation liquid phase, after the fast initial growth,
$\phi$ slowly coarsens due to splitting and further merging of the
domains. This is consistent with \emph{liquid}-like behaviour.
However, a separation of two distinct time scales is clearly
present and we focus on the resulting regime reached after the
fast initial segregation. The data for the microscopic measures
presented below was acquired for a period of 1min and obtained
3min after the driving was switched on.

\begin{figure}
\begin{center}
\includegraphics[width=11cm]{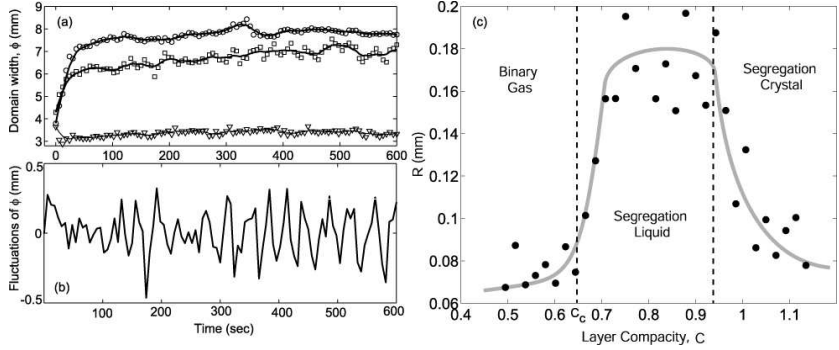}
\caption{(a) Time-series of, $\phi(t)$:  ($\tiny \triangledown$)
for $C=0.580$, ($\tiny \square$) for $C=0.751$ and  ($\circ$) for
$C=1.028$. We filtered $\phi(t)$ through a digital low-pass finite
impulse response (FIR) digital filter algorithm with brickwall at
$\omega=2.29\times10^{-3}Hz$ and $-3dB$ point at
$\omega=9.78\times10^{-3}Hz$ to obtain $\phi_{filtered}(t)$ (solid
curves). (b) Time series of the macroscopic fluctuations,
$f(t)=\phi(t)-\phi_{filtered}(t)$ for $C=0.751$. (c)
Root-mean-square of the fluctuations,  $R$, plotted as a function
of $C$. The dashed line at $C_c=0.647$ denotes the binary gas -
segregation liquid transition point $C_c$ measured in Ref.
\cite{reis:2002}. The dashed line at $C=0.943$ is a guide to the
eye for the possible location of the segregation liquid to crystal
transition. The solid grey line is a guide to the eye. }
\end{center}\end{figure}

Firstly, we  monitor the behaviour of the segregated domains using
a macroscopic measure. We focus on the macroscopic fluctuations
of, $\phi(t)$. This was high pass filtered to remove the slow
dynamics of the domains and thereby obtain their fluctuations,
$f(t)$, directly. An example of $f(t)$ for $C=0.751$ is presented
in Fig. 2b. The fluctuations provide a measure of the collective
noise which arises from the merging and breaking dynamics of the
domains. The RMS of these fluctuations, $R=\sqrt{\langle f(t)^2
\rangle}$, is plotted as a function of $C$ in Fig. 2c. At low
values of $C$ no segregation clusters form and $R$ is low but as
the critical point, $C_c$, is approached, the fluctuations of the
domains grow rapidly as mobile clusters form in the segregation
liquid regime. As $C$ is increased further, $R$ abruptly decreases
when stable crystalline stripes form (see Fig 1d), suggesting a
segregation liquid to crystal transition.

\begin{figure}
\begin{center}
\includegraphics[width=14cm]{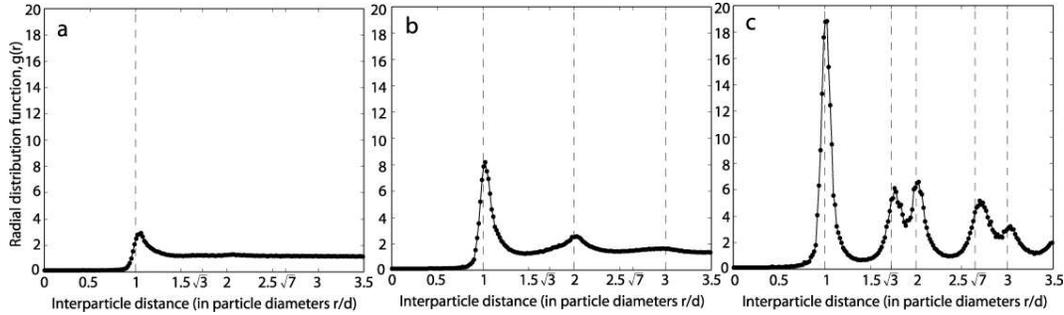} \caption{Radial
distribution functions, $g(r)$, for states in each of the
segregation phases: a) binary gas (C=0.516), b) segregation liquid
(C=0.729) and c) segregation crystal (C=1.071). The vertical
dashed lines correspond to the expected location of the peaks in
each of the segregation phases. Note that the image processing and
finite pixel size of the camera 'blur' the distributions slightly,
which accounts for the non-zero value of $g(r)$, for $r<d$.}
\end{center}\end{figure}

It is now of interest to relate this macroscopic behaviour to the
structural configurations of the spheres as measured by the radial
distribution function,
\begin{equation}
g(r)=A(r)\langle\sum_i\sum_{j\neq i} \delta (r-r_{ij})\rangle,
\end{equation}
where $r_{ij}$ is the separation between the $i$ and $j$th
particles and the angled brackets denote a time average for the
1min of data acquisition. $A(r)$ is a normalization constant such
that $g(r)=1$ for a uniform distribution of particle positions. In
Fig. 3(a,b,c) we show three curves of $g(r)$ corresponding to the
the snapshots (b,c,d) presented in Fig 1, respectively. For the
binary gas (Fig. 3a), $g(r)$ has a peak at $r/d=1$ which quickly
decays at large distances, as expected for a disordered gas.
Increasing $C$ results in a monotonic increase of the height of
the first peak, $g(d)$, consistent with an increasing effective
attractive potential. At intermediate $C$ (Fig. 3b),
\emph{liquid-like} behavior is observed, with $g(r)$ peaked at 1,
2 and 3 particle diameters. The positions of neighbouring spheres
are correlated and the maxima may be associated with concentric
shells of neighbours. The oscillations are rapidly damped, showing
the decay of short-range order. This behaviour is commonly seen in
hard sphere liquids and was first observed experimentally by
Bernal \cite{bernal:1964}. In the segregation crystal phase, at
high $C$, (Fig. 3c) two further peaks are observed near
$r/d=\sqrt{3}$ and $r/d=\sqrt{7}$, characteristic of a static
hexagonally packed crystal, in two dimensions. All peaks are
slightly offset due to the rattling motion of the spheres.

Having presented structural information, we now turn to a
discussion of the local density of the large spheres which was
calculated using Voronoi tessellation \cite{okabe:1992}. The local
density was defined as the ratio between the sphere's projected
area, $\pi (d/2)^2$, and the area of its Voronoi polygonal cell.
This provides a particularly clear measure of the transition
between the binary gas and the segregation liquid phases. We
obtained Probability Distribution Functions (PDF) of the local
density by analyzing 1500 temporal realizations (at $25Hz$),
within the central region where each realization contained $\sim
500$ spheres. A typical example of such a PDF is shown in Fig. 4b.
At a particular $C$, we extracted a characteristic local density,
$\rho_v^{max}$, from the density value at which the peak of the
PDF occurred (Fig. 4c). The width of the PDF, $w$ (Fig. 4c) is a
measure of the local density fluctuations associated with
individual spheres contrasting with the macroscopic fluctuations
which arise from the collective motion of the domains. When the
transition from binary gas to segregation liquid occurred, by
increasing $C$, $\rho_v^{max}$ grew rapidly as segregation domains
formed and there was a significant amplification of the
microscopic density fluctuations at the particular value of
$C_c=0.665\pm0.049$. Close to the segregation transition,
neighboring configurational states coexist and single large
spheres were observed to migrate  between segregation domains.
This rapid growth of an order parameter combined with
amplification of fluctuations is characteristic of equilibrium
phase transitions\cite{binney:1992} giving evidence for the
existence of a critical point for segregation \cite{reis:2002}. A
further indication of critical  behavior is provided by a
time-series of the local density for a typical individual sphere.
Close to this transition point, a large section of the range of
possible local densities ($0<\rho_v\leq\pi/\sqrt{12}$ for disks in
2D) is explored as seen in trace II in Fig. 4a). On the other
hand, a typical time-series in the segregation crystal regime
(trace III in Fig. 4a) illustrates that the particles within the
crystalized segregation domains rattled inside cages formed by
their hexagonally packed neighbors, yielding small fluctuations
about a characteristic high local density which is indicative of a
thermalized crystal.

\begin{figure}[t]
\begin{center}
\includegraphics[width=14cm]{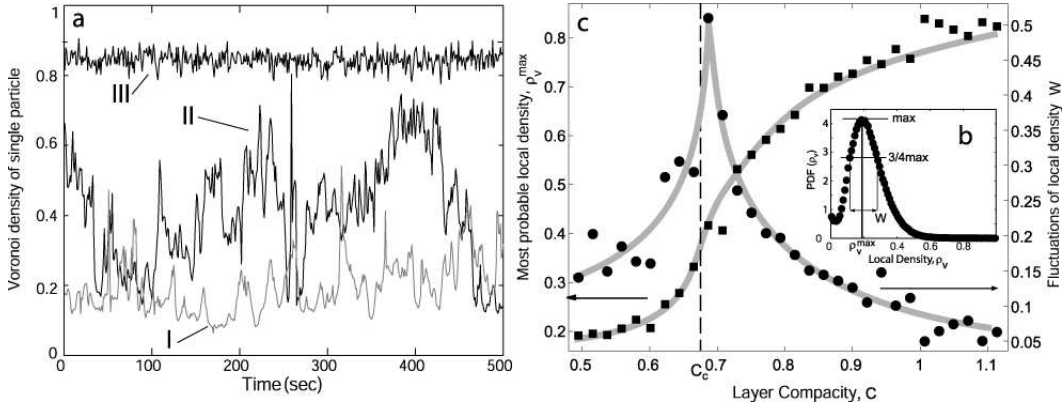}
\caption{Local density. (a) Time series of the local densities for
a typical individual sphere, illustrating the characteristic
fluctuations, for the binary gas ($C=0.578$, trace I), segregation
liquid ($C=0.687$, just above $C_c$, trace II), and segregation
crystal ($C=1.071$, trace III) phases. Inset (b): For each value
of $C$ data, $~750000$ particle positions were used to construct a
PDF for the local density. (c) From this $PDF(\rho_v)$ we extract
the characteristic, i.e. most probable, local density,
$\rho_{v}^{max}$ (squares) and the width gives an estimate of the
microscopic fluctuations, $w$ (circles).  The dashed line is drawn
at the critical compacity value, $C_c$ , measured in
Ref.\cite{reis:2002} and the solid grey lines are guides to the
eye.}
\end{center}\end{figure}

The binary gas to segregation liquid transition occurs at a
critical compacity of $C_c=0.665\pm0.049$. It is interesting to
point out that this value is close to the order-disorder
transition in a two dimensional hard sphere system, which occurs
at a filling fraction of 0.65 for equally sized particles
\cite{luding:2001}.

We note that there are analogies between segregation in our binary
granular system and aggregation in binary colloids
\cite{dinsmore:1995} and emulsions \cite{bibette:1990} where the
ordering is not restricted to differences in size but can also
result from variations in shape or flexibility \cite{adams:1998}.
In particular, when the size ratio between the large colloidal
particles and the radius of gyration of the polymer is
$q\gtrsim0.3$, colloid-polymer mixtures display colloidal gas,
liquid and crystal phases of the large spheres
\cite{poon:2002,poon:2001}. This is consistent with the value of
$q=0.71$ in our granular system.

Aggregation in colloidal systems is usually explained using an
excluded volume depletion argument introduced by Asakura and
Oosawa \cite{asakura:1954,poon:1994} in the context of binary hard
spheres. This entropic argument is equivalent to the mechanistic
view that if two large particles are close enough so that no other
particle (or polymer) may fit between them, they will be subjected
to an asymmetric osmotic pressure that leads to an effective
attractive inter-particle force. Our experimental procedure of
increasing the number of poppy seeds while keeping the number of
spheres constant is analogous to changing the concentration of the
polymer in solution in colloid-polymer mixtures. This deepens the
inter-particle potential $U$ and decreases the overall temperature
$T$ but has the combined effect of increasing the dimensionless
parameter $U/k_BT$ \cite{poon:2002}. The strength of the
interaction can thereby be tuned to induce gas-to-liquid and
liquid-to-crystal transitions \cite{anderson:2002}. While we also
observe qualitatively similar changes in behaviour, our system is
driven, dissipative and far from equilibrium. In addition,
Brownian motion is an intrinsic part of the equilibrium dynamics
of colloidal systems. Hence a direct connection between the two
systems remains speculative.

In conclusion, our results provide quantitative evidence for the
existence of three phases - binary gas, segregation liquid and
segregation crystal - in granular segregation of horizontally
excited binary mixtures. We have presented both macroscopic and
microscopic measures which are self-consistent. Direct analogies
of this phase behaviour may be drawn with other depletion driven
self-assembling binary systems, in particular colloid-polymer
mixtures. This raises the possibility of importing ideas from
binary colloids in equilibrium to formulate new models for
granular segregation.

\acknowledgments

P.M.R. was supported by a scholarship from the Portuguese
Foundation of Science and Technology. G.E. and A.S. were funded by
studentships from the EPSRC. The research of T.M. is supported by
an EPSRC Senior Fellowship. The authors would like to thank D.
Bonamy for advice on the Voronoi density analysis.

\bibliographystyle{unsrt}

\end{document}